\documentclass[reprint,aps,prl,amsmath,amssymb]{revtex4-1}

\usepackage[]{graphicx}
\usepackage[]{units}
\usepackage{times}
\usepackage{bm}
\usepackage{ulem}		
\usepackage{color}

\newcommand{\comment}[1]{}

\renewcommand{\emph}{\textit}

\begin{document}

\title{Ultrafast dynamics in monolayer TMDCs: the interplay of dark excitons, phonons and intervalley Coulomb exchange}
\author{Malte Selig$^1$}
\author{Florian Katsch$^1$}
\author{Robert Schmidt$^2$}
\author{Steffen Michaelis de Vasconcellos$^2$}
\author{Rudolf Bratschitsch$^2$}
\author{Ermin Malic$^3$}
\author{Andreas Knorr$^1$}
\affiliation{$^1$Nichtlineare Optik und Quantenelektronik, Institut f\"ur Theoretische Physik, Technische Universit\"at Berlin,  10623 Berlin, Germany}
\affiliation{$^2$Institute of Physics and Center for Nanotechnology, University of M\"unster, 48149 M\"unster, Germany}
\affiliation{$^3$Chalmers University of Technology, Department of Physics, SE-412 96 Gothenburg, Sweden}

\begin{abstract}

Understanding the ultrafast coupling and relaxation mechanisms between valleys in transition metal dichalcogenide semiconductors is of crucial interest for future valleytronic devices. Recent ultrafast pump-probe experiments showed an unintuitive significant bleaching at the excitonic $B$ transition after optical excitation of the energetically lower excitonic $A$ transition. Here, we present a possible microscopic explanation for this surprising effect. It is based on the joint action of exchange coupling and phonon-mediated thermalization into dark exciton states and does not involve a population of the B exciton. Our work demonstrates how intra- and intervalley coupling on a femtosecond timescale governs the optical valley response of 2D semiconductors.
\end{abstract}

\maketitle

\begin{figure}[t!]
 \begin{center}
\includegraphics[width=0.55\linewidth]{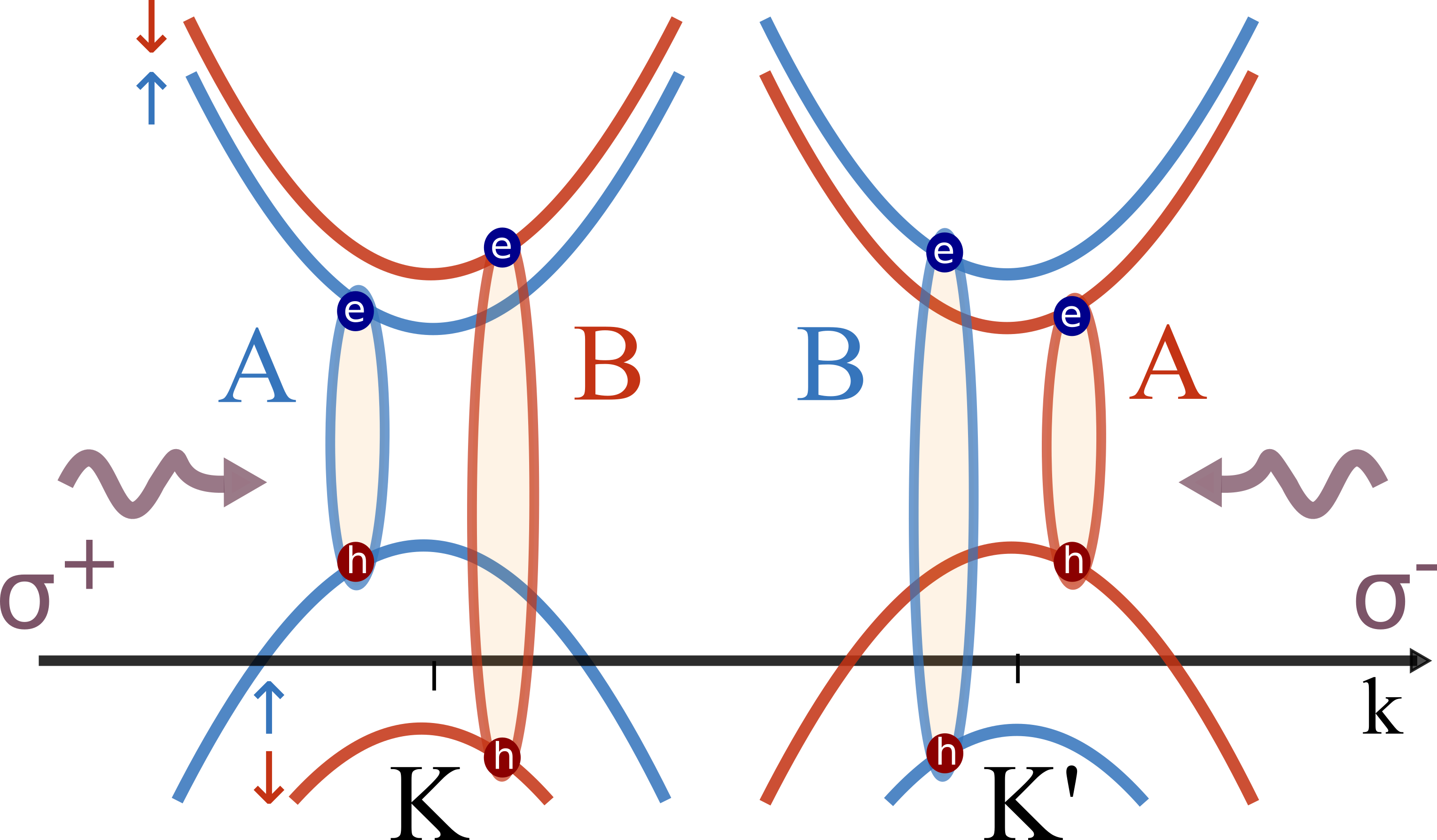}
 \end{center}
 \caption{\textbf{Schematic illustration of the optical selection rules in MoSe$_2$} A pronounced spin-orbit coupling lifts the degeneracy of the electronic bands where the energetic ordering of the different spin bands (blue, red) is reversed between the $K$ to the $K'$ point \cite{Kormanyos2015,Xiao2012}. The spin splitting of the conduction band is on the order of few tens of meV and the splitting of the valence band is on the order of few hundreds of meV. This results in two distinct optical transitions at the $K$ ($K'$) point where the energetically lower (higher) is called the $A$ ($B$) transition \cite{Kormanyos2015}.}
 \label{schema_0}
\end{figure}

\textit{Introduction} Monolayer transition metal dichalcogenides (TMDCs) gained much attention due to their remarkable and rich intra- and intervalley exciton physics.\cite{Berkelbach2013,Li2014,AroraMoSe22015,Qiu2015,Wu2015,Selig2016,Steinhoff2017} The TMDC electronic bandstructure is illustrated in figure \ref{schema_0}: The energetically lowest optically addressable excitons are formed from electrons and holes located at the $K$ and $K'$ points at the corners of the hexagonal Brillouin zone \cite{Steinhoff2014,Kormanyos2015}. These transitions exhibit a circular dichroism i.e. the $K$ ($K'$) point can be excited with left handed (right handed) polarized light $\sigma^+$ ($\sigma^-$) allowing to selectively excite the opposite valleys at the corners of the Brillouin zone \cite{Cao2012}.
On short time scales, polarization ($\sigma^+$, $\sigma^-$) resolved pump-probe signals revealed an ultrafast intervalley coupling finding a sub-picosecond timescale for the intervalley transfer between the pumped and unpumped valley\cite{Wang2013,Conte2015,Schmidt2016,Moody2016,Smolenski2016,Plechinger2016,Tornatzky2018}.

In this Letter, we study the ultrafast intervalley transfer and relaxation dynamics including simultaneous intervalley exchange and phonon scattering and discuss their role for experimentally observed bleaching in transient pump-probe spectra: After optical excitation of the $A$ transition, a significant bleaching at the $B$ transition was observed.\cite{Pogna2016,Bernal2018,Berghauser2018,ZWang2018} Since the energy of the probed $B$ exciton is hundreds of meV larger\cite{Kormanyos2015} compared to the $A$ exciton this type of response comes as a surprise. In particular the helicity resolved data in WS$_2$ show that the bleaching of the B transition occurs not only in the unpumped valley but also within the pumped valley which, without considering further interactions, can only be attributed to an ultrafast spin flip of the electron\cite{ZWang2018} or to a mixture of $A$ and $B$ exciton states\cite{Guo2018}. Interestingly, it was found that the bleaching of the $B$ exciton in the unpumped valley rises faster compared to the pumped valley after optical excitation of the $A$ transition \cite{ZWang2018,Berghauser2018,Bernal2018}.

There are several mechanisms which might explain these results: (a) A spin flip within the pumped valley\cite{Wang2014,Wang2014b} would lead to a population of the opposite-spin conduction band, visible as a $B$ transition bleaching in the same valley\cite{ZWang2018}. (b) Dexter-like intervalley exciton coupling due to electronic wavefunction overlap in k-space where an $A$ exciton in the pumped valley couples directly to the $B$ exciton in the unpumped valley, whereas the $B$ exciton in the pumped valley is driven through intervalley exchange coupling between $B$ excitons\cite{Berghauser2018,Bernal2018}. However, so far, only rough estimates for the coupling strength are avilable\cite{Berghauser2018,Bernal2018}. (c) Beyond intervalley coupling also intravalley exchange coupling leads to an exciton-state mixing between $A$ and $B$ excitons but no polarization selective study of this effect is known yet\cite{Guo2018}. The processes (b) and (c) are strongly off-resonant and a careful evaluation of the actual coupling strength via ab initio methods is required. Another possible explanation for the observed $B$ transition bleaching is, (d) that the $B$ exciton population is driven through stimulated exciton-exciton scattering\cite{Manca2017}. This process is strongly excitation dependent, and therefore contributes only beyond a strict $\chi^{(3)}$ limit.

\begin{figure}[t!]
 \begin{center}
\includegraphics[width=1.0\linewidth]{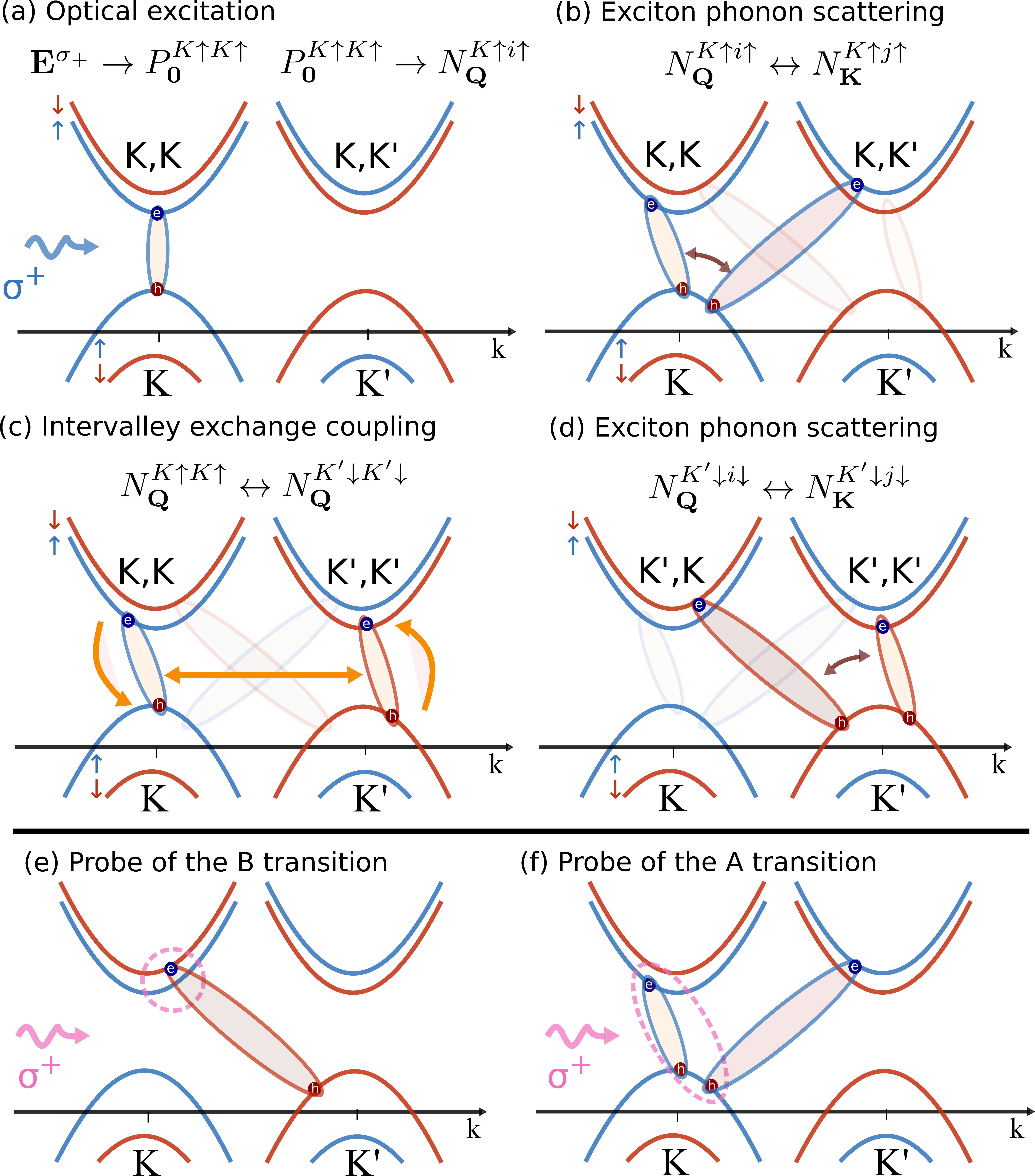}
 \end{center}
 \caption{\textbf{Schematic illustration of the intervalley exchange in MoSe$_2$ (a-d) and the pump-probe experiment (e-f)} (a) Excitation resonantly to the $A$ transition with left handed polarized light $\sigma^+$ (b) Exciton phonon scattering mediates the thermalization of excitons: exciton densities in $(K\uparrow ,K\uparrow )$ states with non-vanishing center of mass momenta are created. (c) Excitons in $(K\uparrow , K\uparrow )$ and $(K'\downarrow ,K'\downarrow )$ states are coupled through intervalley exchange coupling. (d) Exciton phonon scattering between excitonic $\downarrow$ states occurs, which also leads to the population of $(K'\downarrow ,K\downarrow )$ states. (e) Exemplary for the bleaching of the $B$ transition with $\sigma^+$ probe pulse, we find contributions from electron of the $(K'\downarrow , K\downarrow )$ exciton. (f) For the bleaching of the $A$ transition with $\sigma^+$ probe pulse, we find contributions from electron and hole of the $(K\uparrow ,K\uparrow )$ exciton and from the hole of the $(K\uparrow,K'\uparrow )$ exciton.}

 \label{schema}
\end{figure}

All proposed mechanisms (a)-(d) may explain specific features of the polarization resolved ultrafast experiments but so far none of them has explained all related features.
Therefore, in this Letter we introduce a combined mechanism (e), exhibiting several aspects of the mechanisms (a)-(d), at the same time explaining three very specific signatures, seen in all related experiments:

(i) the ultrafast intervalley transfer between the $A$ excitons, 

(ii) the bleaching of the B transition after excitation of the A exciton as well as 

(iii) the temporal ordering of the rise times of the different signals and their temperature dependence. 

Here, we put these observations on a joint consistent theoretical footing: As the dominating mechanism we identify the combined action of the intervalley exchange coupling between the $A$ excitons in both valleys \cite{Maialle1993,Vinattieri1994,Glazov2014,Yu2014,Dery2015,Schmidt2016} and phonon mediated scattering to exciton states consisting of electron and hole at opposite K points in the 1. Brillouin zone\cite{Selig2016,Selig2018} saturating the $B$ exciton transition. Figure \ref{schema} schematically illustrates the proposed mechanism: 

First, figure \ref{schema} (a), an $A$ exciton is excited in the $K$ valley with $\sigma^+$ light. Second, figure \ref{schema} (b), exciton scattering with phonons creates excitons at non-vanishing center of mass momenta, including excitons with electron and hole being located at the same and different high symmetry point. Third, figure \ref{schema} (c), these excitons, having non-vanishing center of mass momenta initialize the intervalley exchange coupling to the unpumped valley. Finally, figure \ref{schema} (d), the excitons in the unpumped valley scatter to intervalley exciton states. 
We propose, that this relaxation dynamics is experimentally visible as a bleaching of the $B$ transition on both valleys ($\sigma^+$, $\sigma^-$), cf. figure \ref{schema} (e), after optical excitation of the $A$ transition in a single valley ($\sigma^+$).

The bleaching results from Pauli blocking due to the cobosonic nature of excitons\cite{Katsch2018}. In particular, for the probe of the $B$ transition, we identify signatures from indirect excitons, where the electron of momentum space indirect excitons occupies the conduction band referring to the $B$ transition leading to a bleaching of the latter in the pump-probe signal, cf. figure \ref{schema} (e). In contrast, for the probe of the $A$ transition we identify contributions from the hole of the momentum indirect excitons and from electron and hole of direct excitons, cf. figure \ref{schema} (f). Interestingly, we find that the response at the $B$ transition in the pumped valley ($\sigma^+$) rises slower compared to the unpumped valley ($\sigma^-$), which results from the step wise process, cf. figure \ref{schema} (a-d), which is required for the formation of the $B$ exciton $\sigma^+$ response. This result is in full agreement with recent experiments\cite{Pogna2016,Schmidt2016,ZWang2018,Berghauser2018}.  
We present detailed numerical calculations for the exemplary material MoSe$_2$, but find that our results are also applicable to other TMDC materials.

\textit{Theoretical Model} We introduce exciton operators $P^{\xi_h \xi_e}_{\mathbf{Q}}$ where $\xi_{e/h} = (i_{e/h} , s_{e/h})$ denotes a compound index consisting of valley and spin of the electron/hole and the excitonic center of mass momentum $\mathbf{Q}$ and an excitonic Hamiltonian\cite{HaugIvanov1993,Katsch2018} to derive the corresponding equations of motion in the Heisenberg picture for expectation values of the excitonic transition $\langle P_\mathbf{Q}^{\xi_h \xi_e} \rangle$ and the incoherent exciton density $N_\mathbf{Q}^{\xi_h \xi_e} = \delta\langle P_\mathbf{Q}^{\dagger\xi_h \xi_e} P_\mathbf{Q}^{\xi_h \xi_e} \rangle$, cf. the supplementary material I. 
The equation of motion for the pumped $A$ exciton transition $\langle P_{\mathbf{Q}=0}^{\xi_h \xi_e} \rangle$ in the $K$ valley (in the incoherent $\chi^{(3)}$ limit \cite{Kochbuch}) reads:
\begin{align}
&i \hbar \partial_t  P^{A \sigma^+}_\mathbf{0} = i \hbar \partial_t  P^{K\uparrow K\uparrow}_\mathbf{0} = \big(E_\mathbf{0}^{K\uparrow K\uparrow}-i \gamma^{K\uparrow K\uparrow}_\mathbf{0}\big)  P^{K\uparrow K\uparrow}_\mathbf{0} \nonumber \\
&+ \mathbf{d}^{K\uparrow} \cdot \mathbf{E}_{pump} \times \nonumber \\ 
&\times \Big( 1 - \sum_{\mathbf{K} i_e} \Xi_{\mathbf{K}}^{K \uparrow , ( K \uparrow i_e\uparrow )} N^{K\uparrow i_e\uparrow}_\mathbf{K} - \Xi_{\mathbf{0}}^{K \uparrow , ( K \uparrow K\uparrow )} |P^{K\uparrow K\uparrow}_\mathbf{0}|^2  \Big) \nonumber \\
&+ \sum_{\mathbf{K},\xi_h,\xi_e} W^{K\uparrow , ( \xi_h,\xi_e)}_\mathbf{K} \big(|P^{\xi_h,\xi_e}_\mathbf{K}|^2  +N^{\xi_h,\xi_e}_\mathbf{K} \big) P^{K\uparrow K\uparrow}_\mathbf{0}.\label{ProbeEqA}
\end{align}

For the $B$ transition in the $K$ valley, tested by the probe pulse, the equation of motion reads:

\begin{align}
&i \hbar \partial_t  P^{B \sigma^+}_\mathbf{0} = i \hbar \partial_t  P^{K\downarrow K\downarrow}_\mathbf{0} = \big(E_\mathbf{0}^{K\downarrow K\downarrow}-i \gamma^{K\downarrow K\downarrow}_\mathbf{0}\big)  P^{K\downarrow K\downarrow}_\mathbf{0}\nonumber \\
 &+ \mathbf{d}^{K\downarrow} \cdot \mathbf{E}_{probe}  \Big( 1 - \sum_{\mathbf{K}} \Xi_{\mathbf{K}}^{ K\downarrow , ( K' \downarrow K \downarrow )} N^{K'\downarrow K\downarrow}_\mathbf{K} \Big) \nonumber \\
&+ \sum_{\mathbf{K},\xi_h,\xi_e} W^{ K\downarrow, ( \xi_h,\xi_e)}_\mathbf{K} \big(|P^{\xi_h,\xi_e}_\mathbf{K}|^2  +N^{\xi_h \xi_e}_\mathbf{K} \big)   P^{K\downarrow K\downarrow}_\mathbf{0}.\label{ProbeEqB}
\end{align} 
The equations of motion for the $A/B$ transition in the $K'$ valley can be obtained by exchanging the high symmetry points $K \leftrightarrow K'$ and spins $\uparrow \leftrightarrow \downarrow$.
In both equations, the first line accounts for the excitonic dispersion with the excitonic energy $E_\mathbf{Q}^{\xi_h \xi_e}$ and the dephasing, where $\gamma^{\xi \xi}_\mathbf{0}$ is determined by radiative decay and exciton phonon coupling \cite{Selig2016,Christiansen2017,Raja2018}. The second line accounts for the optical excitation by the field $\mathbf{E}$ via the dipole moment $\mathbf{d}^{K^{(,)} \uparrow (\downarrow)}$ and for the excitonically modified Pauli blocking of the constituent carriers forming the excitons.
The appearing form factors $\Xi_{\mathbf{K}}^{ \xi , ( \xi_h \xi_e )}$ are a consequence of the co-bosonic commutation relation of excitons\cite{Katsch2018}, equation S5.
Evaluating the appearing sum over the excitonic valley in equation \ref{ProbeEqA} we identify that the bleaching of the $A$ exciton transition results from intravalley $(K,K)$ exciton and the holes of intervalley $(K,K')$ exciton, cf. figure \ref{schema} (f), whereas for the bleaching of the $B$ transition we find contributions from the electrons of intervalley $(K,K')$ excitons only, cf. figure \ref{schema} (e). 
The third line in the equations \ref{ProbeEqA} and \ref{ProbeEqB} schematically describe the energy renormalization due to exciton-exciton interaction with the coupling element $W^{\xi , ( \xi_h \xi_e ) }_\mathbf{K}$, equation S6. Note, that we do not calculate the energy renormalization and further many body effects due to Coulomb interaction, which is also required for the interpretation of pump-probe experiments, as shown in \cite{Pogna2016}, but focus on the bleaching.
The equations \ref{ProbeEqA} and \ref{ProbeEqB} together with the equation of motion of the exciton density, equation S13, could be solved iteratively in orders of the exciting optical field to obtain an analytic expression for the third order susceptibility.


\begin{figure*}[t!]
 \begin{center}
\includegraphics[width=0.9\linewidth]{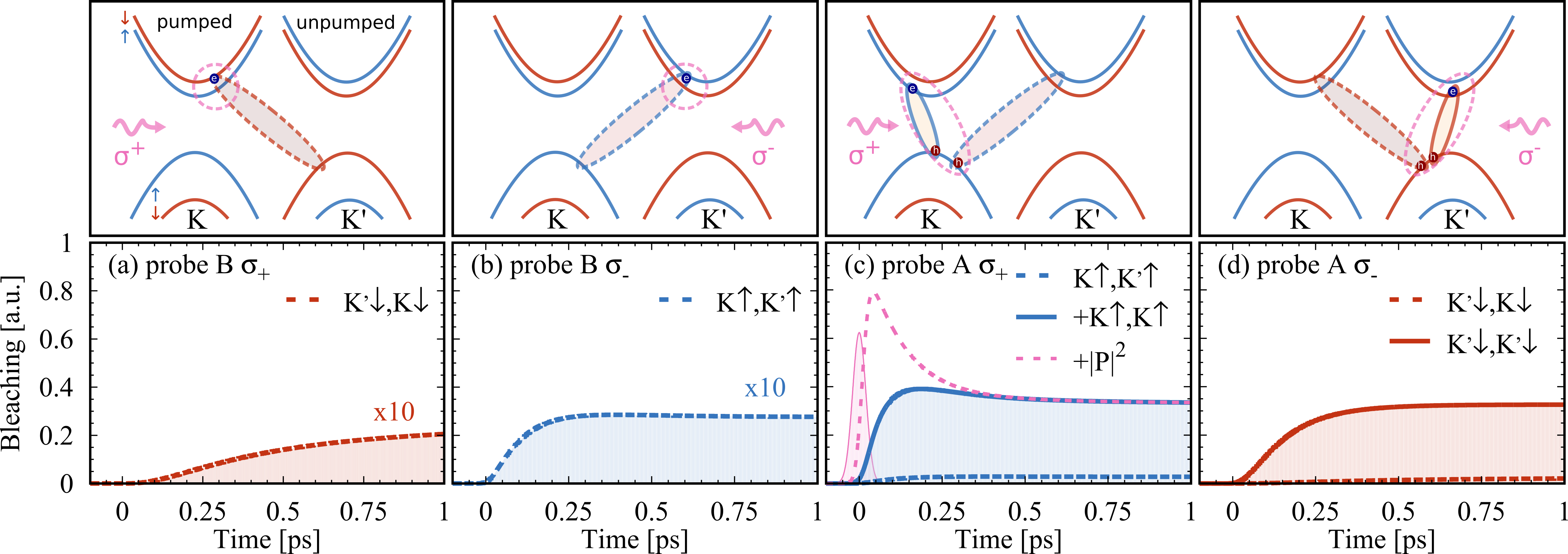}
 \end{center}
 \caption{\textbf{Bleaching of the pump-probe signal at \unit[77]{K} in MoSe$_2$.} Contributions from exciton states with spin $\uparrow$ are denoted in blue, spin $\downarrow$ contributions are denoted in red, as in figure \ref{schema}. In the upper panel, we illustrate the excitons leading to the bleaching of the excitonic transition, which is depicted in the lower panel. Contributions from intravalley excitons are illustrated with solid lines, contributions from intervalley excitons are denoted with dashed lines. (a,b) Pump-probe signal at the $B$ transition, compare equation \ref{ProbeEqB}, in the pumped (a) and unpumped (b) valley after excitation of the $A$ transition. (c,d) Pump-probe signal at the $A$ transition, compare equation \ref{ProbeEqA}, in the pumped (c) and unpumped (d) valley after excitation of the $A$ transition. Additionally we depict the additional bleaching contribution of the optically injected coherence $|P|^2$. The pump pulse with a HWHM of \unit[20]{fs} centered at $t=$\unit[0]{ps} is depicted in figure (c) in pink. }
 \label{PumpProbeMoSe2}
\end{figure*}

\textit{Results} To clarify the temporal pathways of the excitons after optical exciation, we numerically evaluate the coupled dynamics of the excitonic coherence $P_\mathbf{Q}^{K\uparrow K \uparrow}$, the exciton densities $N_\mathbf{Q}^{\xi_h \xi_e}$ and the intervalley coherence $\delta \langle P^{\dagger K \uparrow K \uparrow}_\mathbf{Q} P^{\dagger K' \downarrow K' \downarrow}_\mathbf{Q}  \rangle$ Eqs. S10, S13 and S14. As it is discussed in the supplementary, all parameters are determined by microscopic coupling elements. For our numerical evaluation we explicitly include excitons $( i_h s , i_e s )$ for both spin bands $s = \uparrow,\downarrow$ but identical spin for electron and hole forming the excitons (optically spin allowed) with electron and hole in the same and in different valleys $i_h\neq i_e$. For the hole valleys we include $i_h = K,K'$ and for the electronic valleys we include $i_e=K,K',\Lambda$ (half way between $K$ and $\Gamma$, also referred to as $Q$ or $\Sigma$\cite{Kormanyos2015,Steinhoff2017})$,\Lambda'$. This way, exploiting equations \ref{ProbeEqA} and \ref{ProbeEqB} we have access to the exciton bleaching 
\begin{equation}
B^{\xi} = \sum_{\mathbf{K},\xi_h,\xi_e} \Xi^{\xi,(\xi_h,\xi_e)}_\mathbf{K} N^{\xi_h,\xi_e}_\mathbf{K}.
\end{equation} 
Figure \ref{PumpProbeMoSe2} depicts the polarization ($\sigma^+$, $\sigma^-$) resolved bleaching of the $B$/$A$ excitons seen by a probe pulse (pink wavy line) as a function of time after optical pump with a \unit[20]{fs} $\sigma^+$ pulse resonant to the $A$ transition at an exemplary temperature of \unit[77]{K} in MoSe$_2$. On top figure \ref{PumpProbeMoSe2} additionally illustrates the different bleaching contributions of different intervalley $(K,K')$ and intravalley $(K,K)$ excitons. For all different contributions to the bleaching of $A$ and $B$ excitons we find an ultrafast rise, faster than \unit[1]{ps}: 

Figure \ref{PumpProbeMoSe2} (a) depicts the temporal evolution of the bleaching of the $B$ transition in the pumped valley. As the only contribution we find bleaching from $(K'\downarrow , K\downarrow )$ excitons. The signal rises relatively slow within approximately \unit[700]{fs}, since the formation of these excitons requires a stepwise process as depicted in figure \ref{schema} (a-d): first excitons from the pumped $(K\uparrow , K\uparrow )$ states have to couple to $(K'\downarrow , K'\downarrow )$ exciton states through intravalley phonon scattering (or generate $\mathbf{Q} \neq 0$) and intervalley exchange, cf. figure \ref{schema} (c), and a subsequent phonon scattering to intervalley $(K'\downarrow , K\downarrow )$ states, cf. figure \ref{schema} (d). Such a dynamics is observed as a blocking of the probe pulse and contribute similar to the proposed exciton-upconversion or spin-flips\cite{Manca2017,ZWang2018}. 

Figure \ref{PumpProbeMoSe2} (b) shows the temporal evolution of the bleaching of the $B$ transition in the unpumped valley. The only contribution is the bleaching from $(K\uparrow ,K'\uparrow )$ excitons. The signal rises faster compared to the signal in the pumped valley with a time constant of \unit[120]{fs}, cf. figure \ref{PumpProbeMoSe2} (a), since here the contributing $(K\uparrow ,K'\uparrow )$ excitons can be directly formed through intervalley phonon scattering from the optically pumped $(K\uparrow , K\uparrow )$ excitons, cf. figure \ref{schema} (b). 

Figure \ref{PumpProbeMoSe2} (c) depicts the temporal evolution of the bleaching of the $A$ transition in the pumped valley, where beyond the incoherent excitons $N^{\xi_h \xi_e}_\mathbf{Q}$, also coherent excitons $|P^{K \uparrow K \uparrow}_\mathbf{Q}|^2$ contribute for the resonant probe. As the dominating contribution we identify bleaching from $(K\uparrow , K\uparrow )$ excitons and an additional small contribution from $(K\uparrow , K' \uparrow )$ excitons. The signal rises with the pump pulse, since the $(K \uparrow , K \uparrow )$ excitons are  optically pumped. The additional small contribution from intervalley $(K\uparrow , K' \uparrow )$ excitons is due to phonon mediated thermalization of excitons. Since in MoSe$_2$ these states are located energetically above the bright state by some few meV, their contribution is small at cryogenic temperatures. 

Finally figure \ref{PumpProbeMoSe2} (d) depicts the temporal evolution of the bleaching of the $A$ transition in the unpumped valley. We find dominating contributions from $( K'\downarrow , K'\downarrow )$ excitons and a small contribution from $( K'\downarrow , K\downarrow )$ excitons. The signal rises relatively fast with a time constant of \unit[200]{fs}, since the $( K'\downarrow , K'\downarrow )$ excitons are formed from the pumped $(K \uparrow , K \uparrow )$ excitons through intravalley phonon scattering and intervalley exchange coupling. The small contribution from $( K'\downarrow , K\downarrow )$ excitons is due to the phonon mediated thermalization, as for the pumped valley the $(K'\downarrow , K\downarrow )$ exciton states are located energetically above the $(K'\downarrow , K'\downarrow )$ excitons, explaining their small contribution to the signal. 

All in all, the temporal sequence of the signatures at both $B$ transitions was also observed experimentally recently in WS$_2$\cite{Berghauser2018,ZWang2018}.

Due to the large energetic separation of \unit[150]{meV} from the $A$ exciton in MoSe$_2$\cite{Kormanyos2015}, we find that excitons involving the electronic $\Lambda$ valley have neglectible occupations and a vanishing influence. For the same reason (\unit[180]{meV} detuning) we neglected the lower split off valence band.

\begin{figure}[t!]
 \begin{center}
\includegraphics[width=0.9\linewidth]{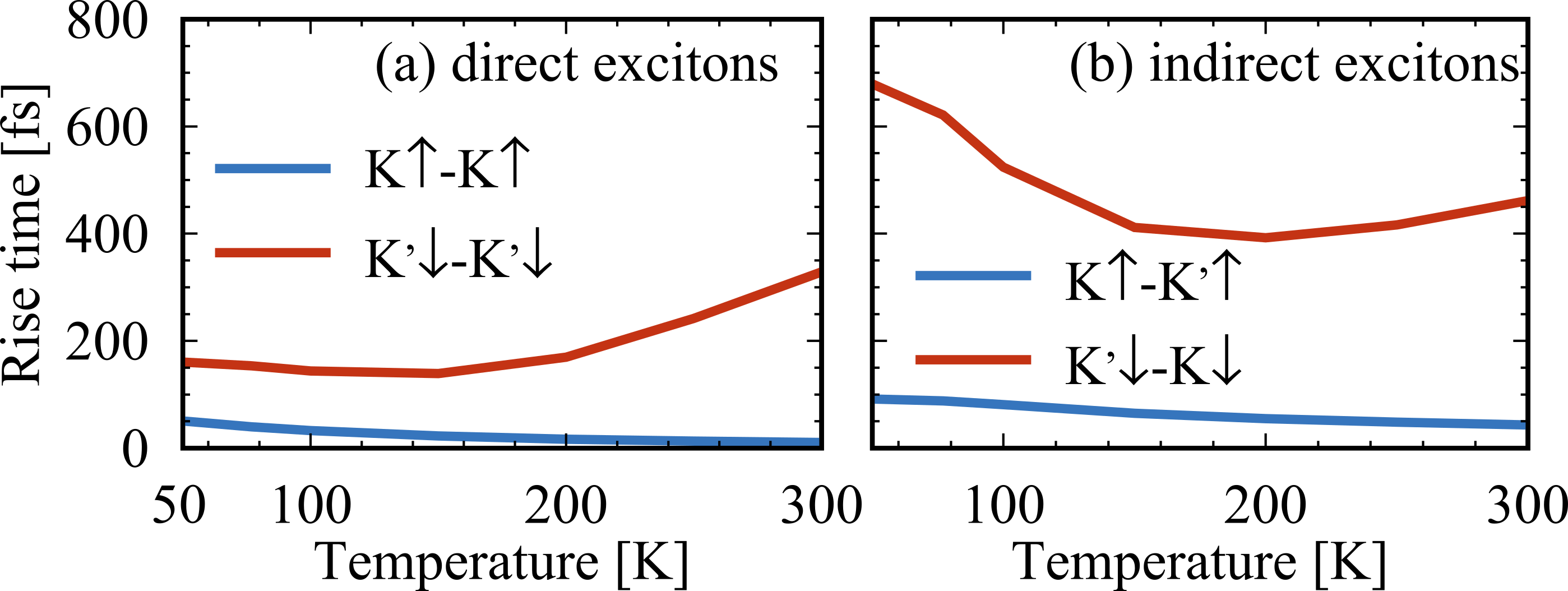}
 \end{center}
 \caption{\textbf{Extracted density rise times in MoSe$_2$.} (a) shows the rise time of the direct exciton densities as a function of temperature and (b) shows the indirect exciton densities. }
 \label{RiseMoSe2}
\end{figure}

To get a more quantitative analysis with respect to experiments\cite{Schmidt2016,Pogna2016,Bernal2018,Berghauser2018,ZWang2018}, we extract the rise times of the wavenumber integrated exciton occupations, c.f. figure \ref{RiseMoSe2}, by fitting their initial temporal evolution exponentially. For the $(K \uparrow ,K \uparrow)$ exciton density, cf. figure \ref{RiseMoSe2} (a), we find decreasing rise times as a function of temperature (about \unit[50]{fs} at \unit[50]{K} and about \unit[10]{fs} at room temperature). This can be understood from the fact, that the $(K \uparrow ,K\uparrow )$ exciton density is directly formed and the optically pumped excitonic coherence fastly converted through intravalley phonon scattering to incoherent excitons\cite{Selig2016,Selig2018}. 
For the $(K' \downarrow ,K'\downarrow )$ density we find an increase from about \unit[160]{fs} at \unit[50]{K} to about \unit[330]{fs} at room temperature. Here, several counteracting effects contribute: At elevated temperatures, since the intervalley exchange coupling increases with exciton momentum, a hotter exciton distribution leads to a faster intervalley transfer.
Counteracting however is the intervalley coherence damping, equation S8, due to phonon scattering at elevated temperatures. Further at elevated temperatures more excitons occupy the indirect $(K \uparrow ,K'\uparrow)$ states, effectively slowing down the intervalley transfer.

In figure \ref{RiseMoSe2} (b) the risetimes of the intervalley exciton occupation is depicted. We find the rise time of the $(K\uparrow , K'\uparrow )$ states decreasing from \unit[90]{fs} at \unit[50]{K} to about \unit[40]{fs} at room temperature, being consistent with the direct formation through phonon scattering from the optically induced exciton coherence. The slowest risetime we find for the $(K'\downarrow ,K\downarrow )$ excitons, ranging from about \unit[680]{fs} at \unit[50]{K} to about \unit[460]{fs} at room temperature. The reason for this comparably slow rise time is that for the formation of these excitons first IEC from the $(K\uparrow ,K\uparrow )$ states to $(K'\downarrow ,K'\downarrow )$ states and after that a phonon mediated scattering from the latter to $(K'\downarrow ,K\downarrow )$ states is required. Our calculated time scale is in good agreement with the experimentally measured rise times of the $B$ signal of about \unit[200]{fs} at \unit[77]{K} in WS$_2$\cite{ZWang2018}. We are aware that in other TMDCs the excitonic landscape deviates from the situation in MoSe$_2$\cite{Selig2018}, and therefore a different temperature dependence can occur. However, we expect, that at least the qualitative behavior including the temporal ordering of the different signals as well as the order of magnitude of the rise times are similar in other materials. Additional to the discussed mechanism also spin-flip processes have been found to contribute to the bleaching of the B transition\cite{ZWang2018}. In our analysis we do not observe any excitation power dependence of the rise times, since we restricted our analysis to the low density limit to keep the already high numerical complexity on a moderate level.

\textit{Conclusion} The microscopic relaxation dynamics including simultaneous intervalley exchange and intervalley phonon scattering significantly contributes to the theoretical understanding
of recent experimental findings, in particular to unintuitive experimental results in helicity-resolved ultrafast pump ($A$ excitons) probe ($B$ excitons) experiments. All presented calculations are consistent with recent experimental observation: (i) the fast transfer between the A excitons due to intervalley
Coulomb exchange \cite{Schmidt2016}, (ii) The bleaching at the $B$ transition after optically exciting the $A$ transition \cite{Pogna2016,Berghauser2018} and (iii) the faster response at the B transition in the unpumped valley compared to the pumped valley\cite{ZWang2018}. However, it does not explain the photoluminescence emission from the B exciton under A exciton excitation\cite{Manca2017}, since no B excitons (involving a hole in the lower valence band) are created.

\begin{acknowledgments}
\textit{Acknowledgements} We acknowledge fruitful discussions with Dominik Christiansen (TU Berlin). This work was funded by the Deutsche Forschungsgemeinschaft (DFG) - Projektnummer 182087777 - SFB
951 (project B12, M.S., A.K.). This project has also received funding from the European Unions Horizon 2020 research and innovation program under Grant Agreement No. 734690 (SONAR, F.K., A.K.). E.M. acknowledges financial support from the European Unions Horizon 2020 research and innovative program under grant agreement No. 696656 (Graphene Flagship) as well as from the Swedish Research Council (VR).
\end{acknowledgments}

\bibliographystyle{naturemag.bst}

\begin{thebibliography}{10}
\expandafter\ifx\csname url\endcsname\relax
  \def\url#1{\texttt{#1}}\fi
\expandafter\ifx\csname urlprefix\endcsname\relax\def\urlprefix{URL }\fi
\providecommand{\bibinfo}[2]{#2}
\providecommand{\eprint}[2][]{\url{#2}}

\bibitem{Berkelbach2013}
\bibinfo{author}{Berkelbach, T.~C.}, \bibinfo{author}{Hybertsen, M.~S.} \&
  \bibinfo{author}{Reichman, D.~R.}
\newblock \bibinfo{title}{Theory of neutral and charged excitons in monolayer
  transition metal dichalcogenides}.
\newblock \emph{\bibinfo{journal}{Phys. Rev. B}} \textbf{\bibinfo{volume}{88}},
  \bibinfo{pages}{045318} (\bibinfo{year}{2013}).

\bibitem{Li2014}
\bibinfo{author}{Li, Y.} \emph{et~al.}
\newblock \bibinfo{title}{Measurement of the optical dielectric function of
  monolayer transition-metal dichalcogenides: $\mathrm{MoS}_{2}$,
  $\mathrm{Mo}\mathrm{S}{\mathrm{e}}_{2}$, $\mathrm{WS}_{2}$, and
  $\mathrm{WS}{\mathrm{e}}_{2}$}.
\newblock \emph{\bibinfo{journal}{Phys. Rev. B}} \textbf{\bibinfo{volume}{90}},
  \bibinfo{pages}{205422} (\bibinfo{year}{2014}).

\bibitem{AroraMoSe22015}
\bibinfo{author}{Arora, A.}, \bibinfo{author}{Nogajewski, K.},
  \bibinfo{author}{Molas, M.}, \bibinfo{author}{Koperski, M.} \&
  \bibinfo{author}{Potemski, M.}
\newblock \bibinfo{title}{Exciton band structure in layered
  $\mathrm{MoSe}_{2}$: from a monolayer to the bulk limit}.
\newblock \emph{\bibinfo{journal}{Nanoscale}} \textbf{\bibinfo{volume}{7}},
  \bibinfo{pages}{20769--20775} (\bibinfo{year}{2015}).

\bibitem{Qiu2015}
\bibinfo{author}{Qiu, D.~Y.}, \bibinfo{author}{Cao, T.} \&
  \bibinfo{author}{Louie, S.~G.}
\newblock \bibinfo{title}{Nonanalyticity, valley quantum phases, and lightlike
  exciton dispersion in monolayer transition metal dichalcogenides: Theory and
  first-principles calculations}.
\newblock \emph{\bibinfo{journal}{Phys. Rev. Lett.}}
  \textbf{\bibinfo{volume}{115}}, \bibinfo{pages}{176801}
  (\bibinfo{year}{2015}).

\bibitem{Wu2015}
\bibinfo{author}{Wu, F.}, \bibinfo{author}{Qu, F.} \&
  \bibinfo{author}{MacDonald, A.~H.}
\newblock \bibinfo{title}{Exciton band structure of monolayer
  ${\mathrm{mos}}_{2}$}.
\newblock \emph{\bibinfo{journal}{Phys. Rev. B}} \textbf{\bibinfo{volume}{91}},
  \bibinfo{pages}{075310} (\bibinfo{year}{2015}).

\bibitem{Selig2016}
\bibinfo{author}{Selig, M.} \emph{et~al.}
\newblock \bibinfo{title}{{Excitonic linewidth and coherence lifetime in
  monolayer transition metal dichalcogenides}}.
\newblock \emph{\bibinfo{journal}{Nature Communications}}
  \textbf{\bibinfo{volume}{7}}, \bibinfo{pages}{13279} (\bibinfo{year}{2016}).

\bibitem{Steinhoff2017}
\bibinfo{author}{Steinhoff, A.} \emph{et~al.}
\newblock \bibinfo{title}{{Exciton fission in monolayer transition metal
  dichalcogenide semiconductors}}.
\newblock \emph{\bibinfo{journal}{Nature Communications}}
  \textbf{\bibinfo{volume}{8}}, \bibinfo{pages}{1166} (\bibinfo{year}{2017}).

\bibitem{Steinhoff2014}
\bibinfo{author}{Steinhoff, A.}, \bibinfo{author}{R\"osner, M.},
  \bibinfo{author}{Jahnke, F.}, \bibinfo{author}{Wehling, T.~O.} \&
  \bibinfo{author}{Gies, C.}
\newblock \bibinfo{title}{Influence of excited carriers on the optical and
  electronic properties of mos2}.
\newblock \emph{\bibinfo{journal}{Nano Letters}} \textbf{\bibinfo{volume}{14}},
  \bibinfo{pages}{3743--3748} (\bibinfo{year}{2014}).

\bibitem{Kormanyos2015}
\bibinfo{author}{Kormanyos, A.} \emph{et~al.}
\newblock \bibinfo{title}{k p theory for two-dimensional transition metal
  dichalcogenide semiconductors}.
\newblock \emph{\bibinfo{journal}{2D Materials}} \textbf{\bibinfo{volume}{2}},
  \bibinfo{pages}{022001} (\bibinfo{year}{2015}).

\bibitem{Cao2012}
\bibinfo{author}{Cao, T.} \emph{et~al.}
\newblock \bibinfo{title}{{Valley-selective circular dichroism of monolayer
  molybdenum disulphide}}.
\newblock \emph{\bibinfo{journal}{Nat Commun}} \textbf{\bibinfo{volume}{3}},
  \bibinfo{pages}{887} (\bibinfo{year}{2012}).

\bibitem{Wang2013}
\bibinfo{author}{Wang, Q.} \emph{et~al.}
\newblock \bibinfo{title}{Valley carrier dynamics in monolayer molybdenum
  disulfide from helicity-resolved ultrafast pump-probe spectroscopy}.
\newblock \emph{\bibinfo{journal}{ACS Nano}} \textbf{\bibinfo{volume}{7}},
  \bibinfo{pages}{11087--11093} (\bibinfo{year}{2013}).

\bibitem{Conte2015}
\bibinfo{author}{Dal~Conte, S.} \emph{et~al.}
\newblock \bibinfo{title}{Ultrafast valley relaxation dynamics in monolayer
  ${\mathrm{mos}}_{2}$ probed by nonequilibrium optical techniques}.
\newblock \emph{\bibinfo{journal}{Phys. Rev. B}} \textbf{\bibinfo{volume}{92}},
  \bibinfo{pages}{235425} (\bibinfo{year}{2015}).

\bibitem{Schmidt2016}
\bibinfo{author}{Schmidt, R.} \emph{et~al.}
\newblock \bibinfo{title}{Ultrafast coulomb-induced intervalley coupling in
  atomically thin ws2}.
\newblock \emph{\bibinfo{journal}{Nano Letters}} \textbf{\bibinfo{volume}{16}},
  \bibinfo{pages}{2945--2950} (\bibinfo{year}{2016}).

\bibitem{Moody2016}
\bibinfo{author}{Moody, G.}, \bibinfo{author}{Schaibley, J.} \&
  \bibinfo{author}{Xu, X.}
\newblock \bibinfo{title}{Exciton dynamics in monolayer transition metal
  dichalcogenides}.
\newblock \emph{\bibinfo{journal}{J. Opt. Soc. Am. B}}
  \textbf{\bibinfo{volume}{33}}, \bibinfo{pages}{C39--C49}
  (\bibinfo{year}{2016}).

\bibitem{Smolenski2016}
\bibinfo{author}{Smole\ifmmode~\acute{n}\else \'{n}\fi{}ski, T.} \emph{et~al.}
\newblock \bibinfo{title}{Tuning valley polarization in a ${\mathrm{wse}}_{2}$
  monolayer with a tiny magnetic field}.
\newblock \emph{\bibinfo{journal}{Phys. Rev. X}} \textbf{\bibinfo{volume}{6}},
  \bibinfo{pages}{021024} (\bibinfo{year}{2016}).

\bibitem{Plechinger2016}
\bibinfo{author}{{Plechinger, Gerd}} \emph{et~al.}
\newblock \bibinfo{title}{{Trion fine structure and coupled spin--valley
  dynamics in monolayer tungsten disulfide}}.
\newblock \emph{\bibinfo{journal}{Nature Communications}}
  \textbf{\bibinfo{volume}{7}}, \bibinfo{pages}{12715} (\bibinfo{year}{2016}).

\bibitem{Tornatzky2018}
\bibinfo{author}{Tornatzky, H.}, \bibinfo{author}{Kaulitz, A.-M.} \&
  \bibinfo{author}{Maultzsch, J.}
\newblock \bibinfo{title}{Resonance profiles of valley polarization in
  single-layer ${\mathrm{mos}}_{2}$ and ${\mathrm{mose}}_{2}$}.
\newblock \emph{\bibinfo{journal}{Phys. Rev. Lett.}}
  \textbf{\bibinfo{volume}{121}}, \bibinfo{pages}{167401}
  (\bibinfo{year}{2018}).

\bibitem{Pogna2016}
\bibinfo{author}{Pogna, E. A.~A.} \emph{et~al.}
\newblock \bibinfo{title}{Photo-induced bandgap renormalization governs the
  ultrafast response of single-layer mos2}.
\newblock \emph{\bibinfo{journal}{ACS Nano}} \textbf{\bibinfo{volume}{10}},
  \bibinfo{pages}{1182--1188} (\bibinfo{year}{2016}).
\newblock \bibinfo{note}{PMID: 26691058}.

\bibitem{Bernal2018}
\bibinfo{author}{Bernal-Villamil, I.} \emph{et~al.}
\newblock \bibinfo{title}{Exciton broadening and band renormalization due to
  dexter-like intervalley coupling}.
\newblock \emph{\bibinfo{journal}{2D Materials}} \textbf{\bibinfo{volume}{5}},
  \bibinfo{pages}{025011} (\bibinfo{year}{2018}).

\bibitem{Berghauser2018}
\bibinfo{author}{{Bergh{\"a}user Gunnar}} \emph{et~al.}
\newblock \bibinfo{title}{{Inverted valley polarization in optically excited
  transition metal dichalcogenides}}.
\newblock \emph{\bibinfo{journal}{Nature Communications}}
  \textbf{\bibinfo{volume}{9}}, \bibinfo{pages}{971} (\bibinfo{year}{2018}).

\bibitem{ZWang2018}
\bibinfo{author}{Wang, Z.} \emph{et~al.}
\newblock \bibinfo{title}{Intravalley spin–flip relaxation dynamics in
  single-layer ws2}.
\newblock \emph{\bibinfo{journal}{Nano Letters}} \textbf{\bibinfo{volume}{18}},
  \bibinfo{pages}{6882--6891} (\bibinfo{year}{2018}).

\bibitem{Guo2018}
\bibinfo{author}{{Guo Liang}} \emph{et~al.}
\newblock \bibinfo{title}{{Exchange-driven intravalley mixing of excitons in
  monolayer transition metal dichalcogenides}}.
\newblock \emph{\bibinfo{journal}{Nature Physics}} \bibinfo{pages}{1745--2481}
  (\bibinfo{year}{2018}).

\bibitem{Manca2017}
\bibinfo{author}{{Manca M.}} \emph{et~al.}
\newblock \bibinfo{title}{{Enabling valley selective exciton scattering in
  monolayer WSe2 through upconversion}}.
\newblock \emph{\bibinfo{journal}{Nature Communications}}
  \textbf{\bibinfo{volume}{8}}, \bibinfo{pages}{14927} (\bibinfo{year}{2017}).

\bibitem{Xiao2012}
\bibinfo{author}{Xiao, D.}, \bibinfo{author}{Liu, G.-B.},
  \bibinfo{author}{Feng, W.}, \bibinfo{author}{Xu, X.} \& \bibinfo{author}{Yao,
  W.}
\newblock \bibinfo{title}{Coupled spin and valley physics in monolayers of
  ${\mathrm{mos}}_{2}$ and other group-vi dichalcogenides}.
\newblock \emph{\bibinfo{journal}{Phys. Rev. Lett.}}
  \textbf{\bibinfo{volume}{108}}, \bibinfo{pages}{196802}
  (\bibinfo{year}{2012}).

\bibitem{Wang2014}
\bibinfo{author}{Wang, L.} \& \bibinfo{author}{Wu, M.}
\newblock \bibinfo{title}{Intrinsic electron spin relaxation due to the
  d'yakonov-perel mechanism in monolayer mos2}.
\newblock \emph{\bibinfo{journal}{Physics Letters A}}
  \textbf{\bibinfo{volume}{378}}, \bibinfo{pages}{1336 -- 1340}
  (\bibinfo{year}{2014}).

\bibitem{Wang2014b}
\bibinfo{author}{Wang, L.} \& \bibinfo{author}{Wu, M.~W.}
\newblock \bibinfo{title}{Electron spin relaxation due to d'yakonov-perel' and
  elliot-yafet mechanisms in monolayer ${\mathrm{mos}}_{2}$: Role of
  intravalley and intervalley processes}.
\newblock \emph{\bibinfo{journal}{Phys. Rev. B}} \textbf{\bibinfo{volume}{89}},
  \bibinfo{pages}{115302} (\bibinfo{year}{2014}).

\bibitem{Maialle1993}
\bibinfo{author}{Maialle, M.~Z.}, \bibinfo{author}{de~Andrada~e Silva, E.~A.}
  \& \bibinfo{author}{Sham, L.~J.}
\newblock \bibinfo{title}{Exciton spin dynamics in quantum wells}.
\newblock \emph{\bibinfo{journal}{Phys. Rev. B}} \textbf{\bibinfo{volume}{47}},
  \bibinfo{pages}{15776--15788} (\bibinfo{year}{1993}).

\bibitem{Vinattieri1994}
\bibinfo{author}{Vinattieri, A.} \emph{et~al.}
\newblock \bibinfo{title}{Exciton dynamics in gaas quantum wells under resonant
  excitation}.
\newblock \emph{\bibinfo{journal}{Phys. Rev. B}} \textbf{\bibinfo{volume}{50}},
  \bibinfo{pages}{10868--10879} (\bibinfo{year}{1994}).

\bibitem{Glazov2014}
\bibinfo{author}{Glazov, M.~M.} \emph{et~al.}
\newblock \bibinfo{title}{Exciton fine structure and spin decoherence in
  monolayers of transition metal dichalcogenides}.
\newblock \emph{\bibinfo{journal}{Phys. Rev. B}} \textbf{\bibinfo{volume}{89}},
  \bibinfo{pages}{201302} (\bibinfo{year}{2014}).

\bibitem{Yu2014}
\bibinfo{author}{Yu, T.} \& \bibinfo{author}{Wu, M.~W.}
\newblock \bibinfo{title}{Valley depolarization due to intervalley and
  intravalley electron-hole exchange interactions in monolayer
  ${\text{mos}}_{2}$}.
\newblock \emph{\bibinfo{journal}{Phys. Rev. B}} \textbf{\bibinfo{volume}{89}},
  \bibinfo{pages}{205303} (\bibinfo{year}{2014}).

\bibitem{Dery2015}
\bibinfo{author}{Dery, H.} \& \bibinfo{author}{Song, Y.}
\newblock \bibinfo{title}{Polarization analysis of excitons in monolayer and
  bilayer transition-metal dichalcogenides}.
\newblock \emph{\bibinfo{journal}{Phys. Rev. B}} \textbf{\bibinfo{volume}{92}},
  \bibinfo{pages}{125431} (\bibinfo{year}{2015}).

\bibitem{Selig2018}
\bibinfo{author}{Selig, M.} \emph{et~al.}
\newblock \bibinfo{title}{Dark and bright exciton formation, thermalization,
  and photoluminescence in monolayer transition metal dichalcogenides}.
\newblock \emph{\bibinfo{journal}{2D Materials}} \textbf{\bibinfo{volume}{5}},
  \bibinfo{pages}{035017} (\bibinfo{year}{2018}).

\bibitem{Katsch2018}
\bibinfo{author}{Katsch, F.}, \bibinfo{author}{Selig, M.},
  \bibinfo{author}{Carmele, A.} \& \bibinfo{author}{Knorr, A.}
\newblock \bibinfo{title}{Theory of exciton-exciton interactions in monolayer
  transition metal dichalcogenides}.
\newblock \emph{\bibinfo{journal}{physica status solidi (b)}}
  \textbf{\bibinfo{volume}{255}}, \bibinfo{pages}{1800185}
  (\bibinfo{year}{2018}).

\bibitem{HaugIvanov1993}
\bibinfo{author}{Ivanov, A.~L.} \& \bibinfo{author}{Haug, H.}
\newblock \bibinfo{title}{Self-consistent theory of the biexciton optical
  nonlinearity}.
\newblock \emph{\bibinfo{journal}{Phys. Rev. B}} \textbf{\bibinfo{volume}{48}},
  \bibinfo{pages}{1490--1504} (\bibinfo{year}{1993}).

\bibitem{Kochbuch}
\bibinfo{author}{Haug, H.} \& \bibinfo{author}{Koch, S.~W.}
\newblock \emph{\bibinfo{title}{Quantum Theory of the Optical and Electronic
  Properties of Semiconductors}} (\bibinfo{publisher}{5th ed. (World Scientific
  Publishing Co. Pre. Ltd., Singapore, 2004).}).

\bibitem{Christiansen2017}
\bibinfo{author}{Christiansen, D.} \emph{et~al.}
\newblock \bibinfo{title}{Phonon sidebands in monolayer transition metal
  dichalcogenides}.
\newblock \emph{\bibinfo{journal}{Phys. Rev. Lett.}}
  \textbf{\bibinfo{volume}{119}}, \bibinfo{pages}{187402}
  (\bibinfo{year}{2017}).

\bibitem{Raja2018}
\bibinfo{author}{Raja, A.} \emph{et~al.}
\newblock \bibinfo{title}{Enhancement of exciton-phonon scattering from
  monolayer to bilayer ws2}.
\newblock \emph{\bibinfo{journal}{Nano Letters}} \textbf{\bibinfo{volume}{18}},
  \bibinfo{pages}{6135--6143} (\bibinfo{year}{2018}).

\end{thebibliography}

\end{document}